# Fabrication and Investigation of Nitrogen doped Ultra-Nano-Crystalline Diamond Hall-bar Devices

N. Eikenberg[1*], K. Ganesan[1], K. K. Lee[1], M. Edmonds[2], L. H. Willems van Beveren[1] and S. Prawer[1]

[1]School of Physics, University of Melbourne, Victoria 3010, Australia
[2]Faculty of Science, Technology and Engineering, La Trobe University, Victoria 3086, Australia
*Corresponding author: Email ninae@student.unimelb.edu.au

**Abstract:** Using microwave-assisted plasma chemical vapour deposition (CVD) a layer of Nitrogen doped ultra-nano-crystalline diamond (N-UNCD) is deposited on top of a non-conducting diamond layer, which itself is situated on a Silicon wafer. This structure is then shaped into Hall-bar devices of various dimensions using optical lithography and dry-etching techniques. The devices' electrical properties are investigated at various temperatures using a cryogen-free dilution refrigerator.

**1 Introduction:** Having many exceptional properties, diamond has long been an attractive material to study. It has a very high thermal conductivity at room temperature, a broad optical transparent window, and it is the hardest material one can find [1]. Through doping the usually insulating material can become a semiconductor and is thus interesting for electronic applications. Boron has successfully been introduced and shown to exhibit superconducting behaviour below a critical temperature of a few Kelvin [2], [3]. Nitrogen can be introduced using a similar CVD process and is another promising dopant, creating free electrons to reveal interesting behaviour at low temperatures [4].

**2 Fabrication:** A Si wafer with a 1 μm thick layer of insulating UNCD is used as a substrate, to ensure that the conducting Silicon base won't interfere with the electrical properties of the N-UNCD. On top of that layer a 1.5 μm thick layer of N-UNCD is created, using microwave-assisted chemical vapour deposition (CVD) with a Nitrogen content of 20% of the gas mixture (Fig. 1 a)). The conditions of the growth are a pressure of 80 Torr, power of 1000 Watt and a plasma temperature of 939 °C.

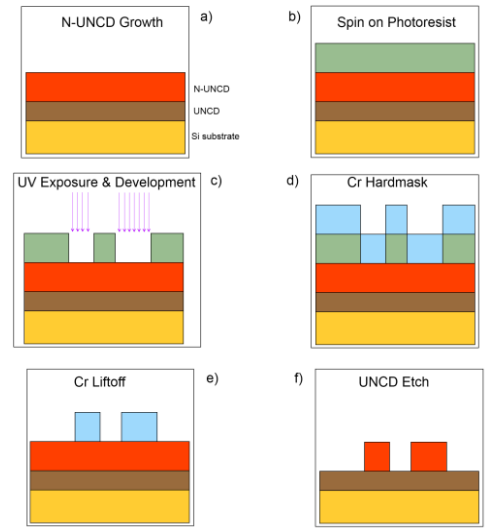

Fig. 1. Sample Fabrication. Panels a) through f) show the steps taken from creating the N-UNCD layer until the Hall-bar-shaped N-UNCD device.

To create the desired structure, optical lithography is used. A positive photoresist is spun onto the sample (Fig. 1 b)) and patterned using a photo mask and exposure to UV light (Fig. 1 c)). After development the structure is visible on the sample. Chromium is evaporated on the whole sample (Fig. 1 d)), covering only the wanted structure after lift-off (Fig. 1 e)). This is used as a hard mask to etch away the remaining diamond so that the result is a device with the desired structure. The etching is done with an inductively coupled plasma – reactive ion etcher (ICP – RIE) using $SF_6$ and $O_2$ gas. After removing the Cr mask through wet etching, a Hall-bar-shaped N-UNCD layer remains (Fig. 1 f)). The six contacts are made accessible by evaporating Ti-Au on top and wire bonding them to a chip carrier.

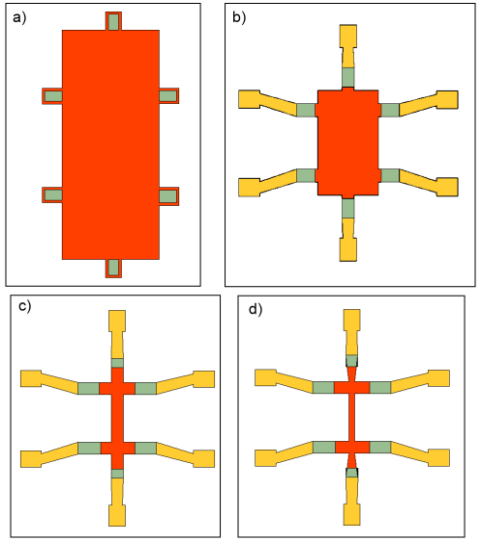

Fig. 2. The four Hall-bar designs. Red represents N-UNCD, yellow the gold contacts, and green is the overlapping area of N-UNCD and gold.

**3 Sample structure:** The chosen sample layout is a Hall-bar design to easily calculate charge carrier mobility and density. Different sizes ensure more accurate results. There are 4 different designs (see Fig. 2) with the following dimensions: a) width (w) = 754 µm, length (l) = 772 µm. b) w = 400 µm, l= 400 µm. c) w = 80 µm, l = 400 µm. d) w = 40 µm, l = 400 µm.

**4 Sample Characterization:** The samples are characterized in a cryogen-free dilution refrigerator at various temperatures, the minimum temperature being 50 mK. For the measurements a variable DC voltage is applied between the top and bottom contacts. Both the longitudinal $V_{xx}$ and the transverse resistance $V_{xy}$ are recorded (Hall-voltage $V_H = V_{xy}$). An external magnetic field is applied perpendicular to the sample surface and varied between 0 and 4 T. This data is then used to calculate the charge carrier density $n$ and mobility $\mu$ using equations (1) and (2):

$$n = \frac{I\,B}{q\,V_H}, \quad (1)$$

where $I$ is the current flowing through the sample, $B$ is the external magnetic field, and $q = -e$ is the carrier charge;

$$\mu = \frac{1}{q\,n\,R_s}, \quad (2)$$

where $R_S$ is the sheet resistivity.

**5 Results:** The measurements are still in progress and the results are therefore not complete. So far our data does not correspond to data reported in the literature [4], as we observe the charge carriers to freeze out below a few Kelvin and data collection becomes challenging.

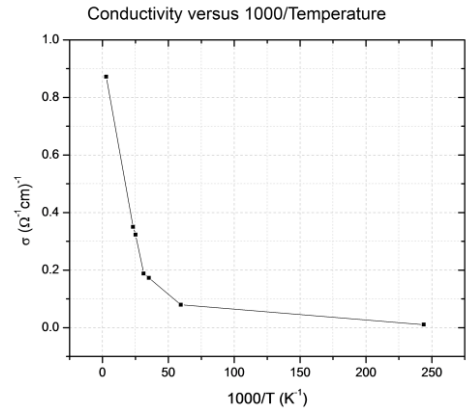

Fig. 3. Conductivity versus 1000/Temperature.

Fig. 3 shows the relationship between the device conductivity and temperature, for a sample with Hall-bar structure (a). The conductivity decreases with reducing temperature as expected. Measurements below 4 K are yet to be performed on this device as well as magnetic field dependant measurements to obtain reliable results ($n$, $\mu$).

**Acknowledgements**
This work was performed in part at the Melbourne Centre for Nanofabrication.